\documentclass[11pt]{article}
\usepackage{amsmath,amssymb,color,epsf,cite}
\usepackage{graphicx}

\makeatletter
\@addtoreset{equation}{section}
\makeatother

\newcommand{\hoch}[1]{$\, ^{#1}$}

\newcommand{\be}{\begin{equation}}
\newcommand{\ee}{\end{equation}}
\newcommand{\bea}{\setlength\arraycolsep{2pt} \begin{eqnarray}}
\newcommand{\eea}{\end{eqnarray}}
\newcommand{\nn}{\nonumber}
\newcommand{\p}{\partial}

\def\ft#1#2{{\textstyle{\frac{\scriptstyle #1}{\scriptstyle #2} } }}
\def\fft#1#2{{\frac{#1}{#2}}}

\def\0{{\sst{(0)}}}
\def\1{{\sst{(1)}}}
\def\2{{\sst{(2)}}}
\def\3{{\sst{(3)}}}
\def\4{{\sst{(4)}}}
\def\5{{\sst{(5)}}}
\def\6{{\sst{(6)}}}
\def\7{{\sst{(7)}}}
\def\8{{\sst{(8)}}}
\def\sst#1{{\scriptscriptstyle #1}}

\def\del{{\partial}}

\def\im{{{\rm i}}}
\def\ie{{i.e.\ }}

\pdfoutput=1
\begin{document}

\begin{flushright}
\hfill{ \
MI-TH-1623\ \ \ \ }
\end{flushright}

\vspace{10pt}
\begin{center}
{\Large {\bf DC Conductivities from Non-Relativistic Scaling Geometries
with Momentum Dissipation}
}

\vspace{30pt}

{\Large
S. Cremonini\hoch{1}, Hai-Shan Liu\hoch{2,3}, H. L\"u\hoch{4}
and C.N. Pope\hoch{3,4,5}
}

\vspace{10pt}

\hoch{1} {\it Department of Physics, Lehigh University,
\it Bethlehem, PA, 18018 USA}

\vspace{10pt}

\hoch{2} {\it Institute for Advanced Physics \& Mathematics,\\
Zhejiang University of Technology, Hangzhou 310023, China}

\vspace{10pt}

\hoch{3} {\it George P. \& Cynthia Woods Mitchell  Institute
for Fundamental Physics and Astronomy,\\
Texas A\&M University, College Station, TX 77843, USA}

\vspace{10pt}

\hoch{4}{\it Center for Advanced Quantum Studies, Department of Physics,\\ Beijing Normal University,
Beijing 100875, China}

\vspace{10pt}

\hoch{5}{\it DAMTP, Centre for Mathematical Sciences,
 Cambridge University,\\  Wilberforce Road, Cambridge CB3 OWA, UK}

\vspace{20pt}

\underline{ABSTRACT}
\end{center}
\vspace{15pt}
We consider a gravitational theory with two Maxwell fields, a dilatonic scalar and spatially dependent axions.
Black brane solutions to this theory are Lifshitz-like and violate hyperscaling.
Working with electrically charged solutions, we calculate analytically the holographic DC conductivities
when both gauge fields are allowed to fluctuate.
We discuss some of the subtleties associated with relating the horizon to the boundary data, focusing on the role of Lifshitz
asymptotics and the presence of multiple gauge fields.
The axionic scalars lead to momentum dissipation in the dual holographic theory.
Finally, we examine the behavior of the DC conductivities as a function of temperature, and
comment on the cases in which one can obtain a linear resistivity.

\thispagestyle{empty}

\pagebreak
\voffset=-40pt
\setcounter{page}{1}

\tableofcontents

\addtocontents{toc}{\protect\setcounter{tocdepth}{2}}

\section{Introduction}

Recent years have seen an increasing interest in applying the techniques of holography to probe
the rich structure of strongly coupled quantum phases of matter, and in particular their dynamics
(see \emph{e.g.} \cite{Hartnoll:2009sz,Sachdev:2011wg,McGreevy:2016myw} for reviews in the context of condensed matter applications).
Efforts are underway to model the transport properties of a variety of systems that exhibit unconventional behavior -- with high temperature
superconductors offering a prime example --
and typically lack a well-defined quasiparticle description, due to their strongly interacting nature.
As part of this program, the breaking of translational invariance (as a mechanism to dissipate momentum
\cite{Karch:2007pd,Hartnoll:2009ns,FILMV,Faulkner:2013bna})
has been recognized as a crucial ingredient for a realistic description
of materials with impurities and an underlying lattice structure (see \emph{e.g.} \cite{Hartnoll:2012rj,Horowitz:2012ky,Horowitz:2012gs,Donos:2012js,
Horowitz:2013jaa,Vegh:2013sk,Davison:2013jba,Chesler:2013qla,Ling:2013nxa,Donos:2014yya,Donos:2013eha, Andrade:2013gsa}).

Indeed, when translational invariance is preserved charges are unable to
dissipate their momentum, and in the presence of
non-zero charge density one encounters a delta function in the AC conductivity
at zero frequency, and a resulting infinite DC conductivity.
Lattice effects and broken translational symmetry have been modeled holographically in a variety of ways.
These include constructions involving periodic potentials and inhomogeneous lattices
\cite{Horowitz:2012ky,Horowitz:2012gs,Horowitz:2013jaa,Chesler:2013qla,Ling:2013nxa,Donos:2014yya},
realizations of homogeneous lattices \cite{Donos:2012js,Donos:2013eha,Andrade:2013gsa} and
theories without diffeomorphism invariance \cite{Vegh:2013sk,Davison:2013jba,Blake:2013bqa}
-- where the list is by no means exhaustive.
The constructions that retain homogeneity involve ordinary (as opposed to partial) differential equations and are therefore of a clear
technical advantage, as they lead to remarkable simplifications in the analysis.

Driven by the desire to model phases with anomalous scalings, there has been
interest in working with geometries that violate hyperscaling
-- describing an anomalous scaling of the free energy parametrized by $\theta$  --
and/or exhibit non-relativistic Lifshitz scaling, characterized by a dynamical critical exponent $z$.
Among the models that maintain homogeneity, conductivity studies for these classes of geometries have appeared in
\cite{Gouteraux:2014hca,Andrade:2016tbr,Amoretti:2016cad,Ge:2016lyn}, with
\cite{Gouteraux:2014hca} focusing on solutions that are
asymptotically AdS.

In this paper, we extend these constructions by examining analytical
black brane backgrounds that are Lifshitz-like and
hyperscaling violating (at all energy scales), and incorporate the
breaking of translational invariance
along the boundary directions by appropriately adding axionic fields.
The theories we consider involve two gauge fields. One is responsible
for the Lifshitz-like nature of the background solutions,
while the other is analogous to a standard Maxwell field in
asymptotically-AdS charged black holes.
Following the horizon method proposed by \cite{Donos:2014uba,Donos:2015gia,Banks:2015wha,Donos:2015bxe}, we
compute analytically the DC conductivities
that encode the response of the system to the presence of the two electric fields.
As we shall see, subtleties arise by taking into account the fluctuations of both fields.

Our analysis complements
the related work of \cite{Gouteraux:2014hca,Lucas:2014sba}, which, however,
considered $\{z,\theta\}$ scaling geometries with AdS UV completions.
In particular, in the appropriate single charge limit, our results provide a concrete realization of one of the IR behaviors
seen in \cite{Gouteraux:2014hca}.
We also examine the structure of the perturbations at the boundary, and
establish how the asymptotic and horizon data are related to each other.
This discussion illustrates how the latter is constrained by choices of boundary conditions, and helps shed some light
on the subtleties associated with working within a non-relativistic theory.

Finally, we shift our attention to the behavior of the DC conductivities
as a function of temperature, restricting ourselves for simplicity
to the regimes that can be treated analytically.
The detailed temperature dependence of holographic conductivities has
received particular attention in the light of the potential
applications to the anomalous ``strange metal'' regime of the high
temperature cuprate superconductors.
A robust feature of the latter is the linear scaling of the resistivity
with temperature, $\rho \sim T$.
With this in mind,
we identify the parameter choices that can lead, in our model, to
a linear resistivity.
In particular, when the only nonzero charge is that of the Lifshitz
gauge field and $z=4/3$,
the leading term (for sufficiently high temperatures)
in the resulting DC conductivity is
$$\sigma_1^{\rm DC}\sim \frac{Q_1^2}{\alpha^2} \frac{1}{T} \, , $$
with $Q_1$ the charge and $\alpha$ the magnitude of the axions -- precisely what one needs to have $\rho \sim T$.
The subleading temperature dependence is then controlled by $\theta$.
We shall find a more intricate temperature behavior when both
charges are turned on, with the existence of a linear regime sensitive
to the sign of the hyperscaling violating exponent $\theta$.

We also study $\sigma_2^{\rm DC}$ as a function of temperature in the cases when $\sigma_1^{\rm DC}=0$.
As a concrete example, in Fig.~1 of section 5 we illustrate some of the possible behavior for the case of $z=4/3$ and $\theta=0$.
At low temperatures, we find a regime in which the resistance $\rho_2=1/\sigma_2$ 
grows linearly with temperature.
At some intermediate temperature $\rho_2$ then turns around and starts decaying, with a behavior that can be approximated \emph{e.g.}
using the Steinhart-Hart equation \cite{steinhart}.

Note that in our setup, since the solutions are asymptotically Lifshitz and
hyperscaling violating,
the DC conductivities will continue to scale even at very large
temperatures, unlike in the case of AdS asymptotics.
Moreover, as we shall see, in order for the background to avoid UV
curvature singularities in the case of Lifshitz asymptotics,
we shall have to impose the condition $\theta>0$, which makes -- in
a number of cases -- the linear part of the resistivity subleading.
Relaxing the condition $\theta>0$ could therefore have interesting
phenomenological consequences.
As an example, the two contributions to
$\sigma_1^{\rm DC}$ both scale as $1/T$
when $z=4/3$ and $\theta=-8/3$, while $z=4/3$ and $\theta=-4/3$ yield
a linear behavior $\rho \sim T$ at large $T$ (in a sense to be made
precise in Section 5), and a quadratic behavior $\rho\sim T^2$ at
smaller temperatures.
Thus, it is important to keep in mind that the same IR features we have
identified here would be present
in models that admit solutions with IR $\{z,\theta\}$ scalings and
AdS UV completions.
Clearly, to construct the latter, the scalar potential of our theory
would need to be appropriately modified, to include terms that would
stabilize the dilatonic scalar in the UV.
While this is certainly possible, by doing so we would lose the advantage
of working with large classes of analytical solutions.
Such an analysis is beyond the scope of the present paper, which focuses
instead on understanding not only the horizon structure of the DC
conductivities,
but also the connection between horizon and boundary data for the case of
Lifshitz asymptotics.

While we were in the final stages of this work, the related article
\cite{Ge:2016lyn} appeared, in which the authors considered the same model
studied here.
However, the analysis of \cite{Ge:2016lyn} only takes into account the
fluctuations of one gauge field.
As we shall explain in detail in the main text, this is not a consistent
truncation of the perturbation equations
-- consistency requires both gauge fields to fluctuate.
This explains the partial discrepancy between our results and those of
\cite{Ge:2016lyn}.

\section{Lifshitz Black Holes with Hyperscaling Violation}

   In this section, we shall consider a particular case amongst the class
of theories described in the appendix, in which we specialise to
four-dimensional gravity coupled
to two Maxwell fields, a dilaton and two axions.  The Lagrangian is
given by
\be
e^{-1} {\cal L} = R - \ft12 (\partial\phi)^2 - 2\Lambda e^{\lambda_0 \phi} -
\ft14 e^{\lambda_1 \phi} F_1^2 - \ft14 e^{\lambda_2 \phi} F_2^2 -
\ft12 e^{\lambda_3 \phi} \big( (\partial \chi_1)^2 +
(\partial\chi_2)^2\big)\,.\label{speclag}
\ee
The equations of motion following from this Lagrangian are
\bea
\label{EOMs}
&&R_{\mu\nu} =  \ft12 \del_\mu\phi\,\del_\nu\phi +
\ft12 e^{\lambda_3\phi}\, (\del_\mu\chi_1\, \del_\nu\chi_1 +\del_\mu\chi_2
\del_\nu\chi_2 ) +\Lambda\, e^{\lambda_0\phi} \, g_{\mu\nu}\nn\\
&&\qquad +\ft12 e^{\lambda_1\phi}\,
 (F_{1\, \mu\rho}\, F_{1\, \nu}{}^\rho -\ft14 F_1^2\, g_{\mu\nu}) +
 \ft12 e^{\lambda_2\phi}\,
 (F_{2\, \mu\rho}\, F_{2\, \nu}{}^\rho -\ft14 F_2^2\, g_{\mu\nu})
\,,\nn\\
&&\square\phi = \ft12 \lambda_3\, \big((\del\chi_1)^2+(\del\chi_2)^2\big)
  +\ft14\lambda_1\,  e^{\lambda_1\phi}\, F_1^2 +
\ft14 \lambda_2\, e^{\lambda_2\phi}\, F_2^2
   +2\lambda_0\, \Lambda\, e^{\lambda_0\phi}\,,\nn\\
&& \nabla_\mu\big( e^{\lambda_3\phi}\, \nabla^\mu\chi_1\big)=0\,,\qquad
\nabla_\mu\big( e^{\lambda_3\phi}\, \nabla^\mu\chi_2\big)=0\,,\nn\\
&& \nabla_\mu\big(e^{\lambda_1\phi}\, F_1^{\mu\nu}\big)=0\,,\qquad
\nabla_\mu\big(e^{\lambda_2\phi}\, F_2^{\mu\nu}\big)=0\,.
\eea

The theory described by (\ref{speclag}) admits Lifshitz-like,
hyperscaling violating black brane solutions, given by\footnote{Our sign convention for $\theta$ is the opposite of the one commonly used
in the literature.}
\bea
ds^2 &=& r^{\theta} \Big(-r^{2z} f dt^2 + \fft{dr^2}{r^2 f} + r^2 (dx^2 +
dy^2)\Big)\,,\cr
\phi &=& \gamma \log r\,,\qquad (A_1)'_0= Q_1\, r^{z-3-\lambda_1 \gamma}\,,\qquad
(A_2)'_0=Q_2\, r^{z-3 - \lambda_2 \gamma}\,,\cr
\chi_1 &=& \alpha x\,,\qquad \chi_2=\alpha y\,,\label{solution}
\eea
and parametrized by
\bea
\label{parameters}
\gamma &=&  \sqrt{(\theta+2)(\theta+2z-2)}\,,\qquad \lambda_0 = -\fft{\theta}{\gamma}\,,\qquad
\lambda_1 = -\fft{(4+\theta)}{\gamma}\,,\cr
\lambda_2 &=& \fft{(\theta + 2z-2)}{\gamma}\,,\qquad \lambda_3 = - \fft{\gamma}{\theta+2}\,,\qquad Q_1 = \sqrt{2(z-1)(\theta+z+2)}\,,\cr
\Lambda &=& - \ft12 (\theta + z+1)(\theta+z+2)\,.
\eea
Note that the logarithmically-running scalar $\phi$
breaks the exact Lifshitz symmetry of the metric.
The blackening function $f$ takes the form
\be
f=1 - \fft{m}{r^{\theta+z+2}} +
\fft{Q_2^2}{2(\theta+2)(\theta+z)\, r^{2(\theta+z+1)}}
+ \fft{\alpha^2}{(\theta+2)(z-2)\, r^{\theta+2z}}\,,\label{fexp}
\ee
with $m$ the integration constant denoting the mass parameter.
The solution is divergent when $z=2$, indicating logarithmic behavior.  Indeed, when $z=2$, the solution becomes
\be
f=1-\fft{1}{r^{\theta+4}}\Big(m + \fft{\alpha^2}{\theta+2}\log r\Big) + \fft{Q_2^2}{2(\theta+2)^2 r^{2(\theta+3)}}\,.
\ee
Thus the $\alpha$ term contributes a logarithmic divergence to the mass.

    For fixed $z$, the solution contains three free integration constants,
$m$, $\alpha$ and $Q_2$.  The solution reduces to the Lifshitz-like vacuum when these parameters vanish.
In this paper, we are not only considering the IR region near the
black hole horizon, but the entire black hole solution (\ref{solution})
including its
asymptotic properties at infinity.
In order for the vacuum to avoid a curvature singularity at the asymptotic
boundary
$r=\infty$, we must require\footnote{If one treats the solution
(\ref{solution}) as
merely an approximation to the geometry in the IR region, this condition can be relaxed, and negative values of $\theta$ are allowed.}
\be
\theta \ge 0\,.\label{thetapositive}
\ee
We must also require
\be
(2+\theta)(2z-2+\theta)\geq 0 \,,\qquad
(z-1)(2+z+\theta) \geq 0 \,,
\ee
in order to ensure that $\gamma$ and $Q_1$ are real.  These last two conditions
are in fact equivalent to the requirement that the null energy condition be
satisfied.  Thus, since $z$ must necessarily be positive,
we must have $z\ge 1$.  This, together with (\ref{thetapositive}),
implies $f(\infty)=1$.  Thus the solution is asymptotically Lifshitz-like
with hyperscaling violation.  In fact the solution (\ref{solution})
describes a charged black hole whose
Hawking temperature is
\be
T= \fft{r_0^{z+1}\, f'(r_0)}{4\pi}\,,\label{hawking}
\ee
where $r_0$ is the radius of the horizon, located at the largest root
of $f(r)=0$.

  It should be noted that the two Maxwell fields play very different roles
in the black hole solution.  The field $A_1$ is responsible for
the Lifshitz-like nature of the vacuum.
In particular, its ``charge''
$Q_1$ is fixed, for given Lifshitz and hyperscaling violating exponents $z$ and
$\theta$, and the solution becomes asymptotically AdS if $Q_1=0$.  By contrast,
the charge $Q_2$ of the field $A_2$ is a freely-specifiable parameter,
analogous to the electric charge of a Reissner-Nordstr\"om black hole.

For reasons that will become apparent shortly, and to make contact
with some of the literature, we would like to
express the scaling of the gauge field responsible for sourcing the
Lifshitz background in terms of the conduction exponent $\zeta$ that controls
the anomalous scaling dimension of the charge density operator\cite{Gouteraux:2012yr,Gouteraux:2013oca,Karch:2014mba}.
Letting\footnote{Note that $\zeta_1=-d_\theta$, where
$d_\theta \equiv 2 + \theta$ is the effective dimensionality factor in four space-time dimensions.}
$\zeta_1 = -2 -\theta$, the scaling of the $A_1$ gauge field is then of the form
\be
\label{zeta1}
A_1 \sim r^{z-\zeta_1} dt \, .
\ee
Similarly, we can introduce another parameter $\zeta_2$, so that the second gauge field can be written as
\be
\label{zeta2}
A_2 \sim r^{z-\zeta_2} dt \, ,
\ee
where we now have $\zeta_2 = 2z+\theta$.

\section{DC Conductivity from Horizon Data}
\label{DCansatzsec}

By now there are several techniques available for computing
holographic DC conductivities.
Using Kubo's formula, the optical conductivity can be extracted from the current-current propagator in the boundary,
\be
\label{Kubo}
\sigma^{ij}(\omega) = \frac{\p}{\p E_j(\omega)} \langle J^i(\omega) \rangle
= -\frac{1}{{\rm i} \omega} \langle J^i(\omega) J^j (\omega) \rangle \, ,
\ee
with the current found by varying the action with respect to the
external source, \emph{i.e.} schematically $\langle J(\omega) \rangle  =
\frac{\p S}{\p A_{\rm ext}(\omega)} $.
The DC conductivity is then simply the zero frequency limit of the optical conductivity,
\be
\sigma^{ij}_{\rm DC} = \lim_{\omega \rightarrow 0} \sigma^{ij}(\omega) \, .
\ee
A great simplification in these calculations comes from the membrane paradigm approach of \cite{Iqbal:2008by},
\emph{i.e.} the realization that the currents in the boundary theory can be identified with
radially independent quantities in the bulk.
In the presence of momentum dissipation, the method of \cite{Iqbal:2008by} was first extended by \cite{Blake:2013bqa}, who noted that
 one can generically identify -- in the zero frequency limit -- a massless mode that does not evolve between the horizon and the boundary.
A much more general understanding of this behavior, and in particular of the universality of the
equivalence between horizon and boundary current fluxes, was later obtained in
\cite{Donos:2014uba,Donos:2015gia,Banks:2015wha,Donos:2015bxe}. Moreover, it was shown \cite{Donos:2015gia,Banks:2015wha,Donos:2015bxe}
that the field theory thermoelectric DC conductivity can be found by solving
generalized Stokes equations on the black hole horizon.

The general procedure for computing DC conductivities entails
studying time-dependent perturbations of the relevant fields.
In particular, one typically turns on an electric field (proportional to $e^{-\im \omega t}$) with frequency $\omega$,
computes the response and then takes the $\omega \rightarrow 0$ limit to extract $\sigma_{\rm DC}$.
However, in the newer approach developed by \cite{Donos:2014uba},
instead of taking the zero-frequency limit of the optical
conductivity, one switches on a constant electric field from the
start\footnote{This amounts to just considering the first two terms
in the Taylor expansion of $e^{-\im \omega t}$, that is to say, considering
perturbations that have terms independent of $t$ and terms linear in $t$.
In fact, the only terms where linear $t$ dependence arises are in the
perturbations of the gauge field potentials: The associated
static electric fields are described in terms of gauge potentials
that depend linearly on $t$.}, and then computes the response.
In this section we will adopt the horizon approach of \cite{Donos:2014uba,Donos:2015gia,Banks:2015wha,Donos:2015bxe}.

   Following the ansatz of \cite{Donos:2014uba},
   we therefore consider perturbations
\be
(\delta A_i)_x =-E_i t + \hat a_i (r)\,,\qquad
\delta g_{tx}  = r^{\theta+2}\, \hat\psi(r)\,,\qquad \delta \chi_1=\hat b(r)\,,
\label{DCperts}
\ee
where hatted quantities are introduced to distinguish the fluctuations
from those of the next section.
Note that in the literature, a $\delta g_{rx}$ perturbation is sometimes
included.  This, however, is pure gauge, and can be removed by an appropriate
coordinate transformation $x\rightarrow x + \beta(r)$, together with a
corresponding field redefinition of $\hat b(r)$.

   Substituting (\ref{DCperts}) into the equations of motion (\ref{EOMs}),
the two Maxwell equations imply $j_1'=0$ and $j_2'=0$, where
\be
j_1=-\big(Q_1 \hat\psi + r^{z-3-\theta} f\hat a_1'\big)\,,\qquad
j_2=-\big(Q_2 \hat\psi + r^{3z-1+\theta} f\hat a_2'\big)\,.\label{hypj1j2}
\ee
Thus $j_1$ and $j_2$ are constants of integration.
They of course describe precisely the two conserved currents in the system.
Similarly, the  axion equations imply $j_0'=0$, where
\be
j_0 = r^{5-z} f\, \hat b'\,,
\ee
and hence $j_0$ is another constant of integration.
The Einstein equations then imply
\be
E_1 Q_1 + E_2 Q_2 = j_0 \alpha\,,
\ee
and
\be
\Big(r^{5-z+\theta}\, \hat\psi'+ Q_1 \hat a_1 + Q_2 \hat a_2\Big)'
  = \fft{\alpha^2}{r^{3z-3}\, f}\, \hat\psi\,.
\ee
Using (\ref{hypj1j2}), this becomes
\be
f\, (r^{5-z+\theta} \hat\psi')' = \fft{1}{r^{3z-1+\theta}}\Big[
\left(Q_2^2 + r^{2+\theta} (\alpha^2 + Q_1^2 r^{(2z+\theta)})\right)\hat\psi +
j_2 Q_2 + j_1 Q_1 r^{2(z+1 + \theta)}\Big]\,.\label{psipp}
\ee
Finally, the dilaton equation gives no contribution at linear order
in perturbations.

   As we mentioned earlier, the field $A_2$ is analogous to a
standard Maxwell field in an asymptotically-AdS charged black hole,
whereas the field $A_1$ is responsible for modifying the vacuum to
become Lifshitz-like.  It is thus tempting to think that one could
consider perturbations around the background in which only
$A_2$, but not $A_1$, is allowed to fluctuate.  In fact, this is what
was done in references \cite{colgain,Ge:2016lyn}.
However, as can be seen from
eqns (\ref{hypj1j2}), turning off the perturbation $\hat a_1$ forces $\hat\psi$
to be a constant, which from (\ref{psipp}) implies that $j_1 Q_2=j_2 Q_1$
and (if $\alpha^2\ne0$) that $j_1=0$.
In all cases, the equation for
$\hat a_2$ in (\ref{hypj1j2}) then implies that $\hat a_2$ is a constant
(which means it is pure gauge).  In other words, a truncation where the
perturbation of $\hat a_1$ is set to zero is inconsistent with the
full set of equations of motion.  In our discussion, we shall therefore
take $\hat a_1$, as well as $\hat a_2$, to be non-vanishing.

   The two leading-order terms in the large-$r$ expansion of
 $\psi$ at large $r$ can be seen, from (\ref{psipp}), to be of the form
\be
\hat\psi = - \fft{j_1}{Q_1} +
  \fft{\beta_1}{r^{z+2+\theta}} + \beta_2\, r^{2z-2} +
\cdots\,.\label{hatpsiinfinity}
\ee
We must take $\beta_2=0$ for regularity, and hence we have
\be
\hat\psi_\infty=-\fft{j_1}{Q_1}\,,
\ee
where $\hat\psi_\infty\equiv \hat\psi(\infty)$.
Evaluating (\ref{psipp}) on the horizon implies that
\be
\label{psi0hat}
\hat\psi_0 = -\fft{j_2\, Q_2 + j_1 \, Q_1 \, r_0^{2(z+1+\theta)}}{Q_2^2 +
(\alpha^2 + Q_1^2 \, r_0^{2z+\theta})\,  r_0^{2+\theta}}\,,
\ee
where $\hat\psi_0\equiv \hat\psi(r_0)$. Now, it follows from (\ref{DCperts})
that in order for the perturbations $(\delta A_i)_x$ to be purely ingoing
on the horizon, we must have $\hat a_i\sim -E_i\, r_*$ near the horizon,
where the tortoise coordinate $r_*$ is defined by $dr_* = dr/(r^{z+1}\, f(r))$.
Thus near the horizon we must have
\be
\hat a_i' = -\fft{E_i}{r^{z+1}\, f(r)} +\cdots \,,
\ee
and so, from (\ref{hypj1j2}), we have
\be
j_1=-Q_1\hat\psi_0 + \fft{E_1}{r_0^{4+\theta}}\,,\qquad
j_2=-Q_2 \hat\psi_0 + E_2 r_0^{2z-2+\theta}\,.
\ee
Finally, combining these with (\ref{psi0hat}) and expressing $j_1,j_2$ entirely in terms of $E_1$ and $E_2$,
in analogy with Ohm's law, we find
the following matrix-valued equation for the currents,
\be
\label{condmatrix}
\left(
\begin{array}{c}
j_1 \\ j_2
\end{array}
 \right)  =
\left(
  \begin{array}{cc}
    \sigma_{11} & \sigma_{12} \\
    \sigma_{21} & \sigma_{22} \\
  \end{array}
\right)
\left(
\begin{array}{c}
E_1 \\ E_2
\end{array}
 \right)\,,
\ee
with the entries of the conductivity matrix given by
\bea
\sigma_{11} &=&\fft{1}{r_0^{4+\theta}} + \fft{Q_1^2 \,r_0^{2z-4}}{\alpha^2}\,,
\qquad
\sigma_{12}=\fft{Q_1Q_2r_0^{2z-4}}{\alpha^2}\,,\cr
\sigma_{21}&=&\sigma_{12}\,,\qquad
\sigma_{22}=r_0^{2z-2+\theta} +
\fft{Q_2^2}{\alpha^2} r_0^{2z-4}\,.\label{sigcomps}
\eea
Recall, however, that here the two currents $j_1$ and $j_2$ are associated with two distinct electric fields,
which are oriented along the same spatial direction. Thus, the
conductivity coefficients appearing in the matrix
should not be confused with those associated with
different space-time directions.
At this point we should note that although $\sigma_{22}$ above
naively agrees with the result of \cite{Ge:2016lyn},
the latter did not take into account both gauge field fluctuations.
Indeed, setting $Q_1=0$ in the equations above is \emph{not} consistent
with the linearised fluctuation equations unless one also sets $z=1$.

As already seen in a number of models in the literature, there are the two types of contributions to the conductivity matrix (\ref{sigcomps}),
only one of which depends explicitly on the magnitude of the axionic sources and the charge(s).
How the dissipative and non-dissipative effects arrange themselves to form the DC conductivity has been
understood \cite{Davison:2015bea} for relativistic theories (in the limit of weak momentum relaxation), \emph{i.e.} backgrounds with AdS asymptotics.
However, to the best of our knowledge in Lifshitz backgrounds this is not the case, and clarifying
how the physics of momentum dissipation is encoded into the final result for $\sigma_{\rm DC}$ is an interesting question which requires
carefully examining the nature of transport in non-relativistic theories.

Also, the fact that $\sigma_{12}$ scales just like the $\alpha$-dependent parts of $\sigma_{11}$ and $\sigma_{22}$
is analogous to what was observed in \cite{Blake:2014yla}, although in a different context
(the focus of \cite{Blake:2014yla} was the temperature dependence of the Hall angle).
While we are not including here the effects of a magnetic field -- we leave the analysis of the Hall angle for future work -- we expect to
see the same generic behavior observed in \cite{Blake:2014yla},
with the difference between the scalings of the diagonal and off-diagonal terms simply due to the presence of
different scales in the system. 

Let us first discuss the single charge case, where we set $Q_2=0$.
The matrix $\sigma_{ij}$ in  (\ref{condmatrix}) then
becomes diagonal.
One can further truncate out the gauge field $A_2$
from the theory.  The resulting conductivity for the remaining
gauge field $A_1$ is then simply $\sigma^{\rm DC}_1=\sigma_{11}$.
Written in terms of the scalar couplings appearing in the Lagrangian,
we have
\be
\sigma^{\rm DC}_1 = e^{\lambda_1 \phi_0} +
   \frac{Q_1^2}{\alpha^2\, r_0^{2+\theta}} \, e^{-\lambda_3 \phi_0} \,,
\label{goutDC}
\ee
which makes it apparent that the coupling between the axionic fields and the dilatonic scalar
is responsible for generating additional temperature dependent terms,
sensitive to the mechanism to relax momentum.
In particular, as we shall see in Section 5, the term proportional to $Q_1^2/\alpha^2$ can
give a temperature dependence of the form $\sigma^{\rm DC}_1\sim 1/T$,
and therefore a linear behavior for the resistivity.
Finally, we should note that (\ref{goutDC}) is precisely of the form
found in \cite{Gouteraux:2014hca},
whose model matches ours when $Q_2=0$.
In the setup of \cite{Gouteraux:2014hca}, however, the background solutions
are assumed to have AdS asymptotics, and the focus is on the possible
IR behavior of the geometry, which is taken to be of the hyperscaling
violating and Lifshitz form.
We shall return to the issue of boundary conditions and asymptotics
in the next section.

It is also worth noting that in the limit $z\rightarrow 1$ and
$\theta\rightarrow 0$, for which $Q_1\rightarrow 0$, the quantity
$\sigma_{11}$ goes to $r_0^{-4}$.  On the other hand, if we had set
$Q_1=0$ from the outset, the solution would have been simply
the Schwarzschild-AdS black hole, and hence we would have $\sigma_{11}=1$.
Thus we have a discontinuity in the $Q_1\rightarrow0$ limit.  This
discontinuity can be understood from the fact that turning on $Q_1$
changes the asymptotic structure from an AdS to a Lifshitz like one.  No
matter how small $Q_1$ is, the perturbation $(\delta A_1)_x$ of the
associated gauge potential must be even smaller.  Thus the perturbation we
are considering here would actually vanish in the $Q_1\rightarrow 0$
limit, and so it could not possibly have a continuous
limit  to the perturbation that is
normally considered in the $Q_1=0$ Schwarzschild-AdS background.
This may be why the dependence of the conductivity on
the temperature is different from that in the RN black hole.

Next, we would like to examine the special case for which $j_1=0$,
\emph{i.e.} the current
associated with fluctuations of the gauge field responsible for the
Lifshitz background vanishes.  Note that this does not
mean that the fluctuation $\delta(A_1)_x$ is turned off.
As we shall see in the next section, the regularity of the perturbation
at infinity requires $j_1=0$ for $1\le z \le \ft43$.
This then describes the case in which the charged degrees of freedom
associated  with $A_1$ are insulating.  For $z>\ft43$, it is not
obligatory to set $j_1=0$.
We can use (\ref{condmatrix}) with $j_1=0$ to trade $E_1$ for $E_2$, and
extract the DC conductivity by reorganizing the resulting expression
for the remaining current, $j_2 = \sigma^{\rm DC}_2 \, E_2$. We find
\be
\label{jzerosigmaDC}
\sigma_2^{\rm DC} = r_0^{2z-2+\theta}\, \left[1 +
\fft{Q_2^2}{r_0^{2+\theta} (\alpha^2 + Q_1^2 \, r_0^{2z+\theta})}\right]\,.
\ee
We emphasize that this is \emph{not} the same as the original $\sigma_{22}$.
Indeed, this expression is sensitive to the presence of both charges.
In particular, the contribution from $Q_1$ introduces additional
temperature dependence, which is absent in $\sigma_{22}$ and would also
not be present in the single charge case above (for which $Q_2=0$).
This additional dependence was also missed by \cite{Ge:2016lyn}, as we
emphasized earlier.

Note that when  $Q_1=0$ we lose the Lifshitz scaling of the
background ($z=1$).
When both charges vanish, the DC conductivity (\ref{jzerosigmaDC}) becomes
\be
\sigma^{\rm DC}_2 = r_0^\theta \,,
\ee
corresponding to a geometry that is conformal to AdS, and reduces to the
well-known result of \cite{Herzog:2007ij},  $\sigma^{\rm DC}_2=1$,
when the background respects hyperscaling, $\theta=0$.
Thus we see that the gauge field $A_2$ and its charge $Q_2$ can be viewed
as generalisations of the gauge field in the Reissner-Nordstr\"om
black hole.  In particular, as we will see shortly, $\sigma_2^{\rm DC}$ cannot have
a $1/T$ temperature dependence
when $\theta>0$, at least at large temperatures.
We shall discuss this in greater detail in
section \ref{Tdep}, where we shall examine the specific temperature
dependence of the DC conductivities.

  Finally, it is worth remarking that the conductivities we obtained in this
section were solely obtained from the horizon data.  The requirement that
the perturbations be well behaved at infinity may give further constraints
on the parameters in the solution.  Since the solutions
(\ref{solution}) we are considering allow us to study the perturbations in the
entire region exterior to the black hole, we will indeed obtain such constraints from inspecting the
asymptotic behaviour of the fluctuations, as we discuss next.

\section{Asymptotic Analysis}
\label{monosec}

We are now ready to turn our attention to the asymptotic analysis, which
will offer an alternative way to obtain the
conductivity matrix we have just derived using the horizon method.
As we shall see, the presence of multiple gauge fields substantially
complicates the analysis.
Here we will highlight some of the subtleties which arise from allowing
each gauge field to fluctuate, and comment on how this method
relates to the one of Section \ref{DCansatzsec}.

In contrast to the analysis of Section 3 -- in which the electric
fields were taken to be constant -- we will now allow
the fluctuations of all the fields
to have monochromatic time dependence $e^{-\im\omega t}$.
To this end, we consider the following perturbations:
\be
(\delta A_i)_{x_1} = a_i(r) e^{-{\rm i}\omega t}\,,\qquad
\delta \chi_1 = b(r) e^{-{\rm i}\omega t}\,,\qquad
\delta g_{tx} = r^{\theta+2}\,
    \psi(r) e^{-{\rm i}\omega t}\,,\label{complexrhs}
\ee
where it is to be understood that the physical perturbations are given by
taking the real parts of these expressions.
At the linear level, the equations of motion then imply
\bea
&&(r^{z-3-\theta} f a_1')' + \fft{\omega^2 a_1}{r^{z+5+\theta}f} +
Q_1 \psi'=0\,,\nn\\
&&(r^{3z-1+\theta} f a_2')' + \fft{\omega^2 a_2}{r^{3-z-\theta}f} +
Q_2 \psi'=0\,,\nn\\
&&\psi' = -\fft{1}{r^{5-z + \theta}}\Big( Q_1 a_1 + Q_2 a_2
- \fft{r^{5-z}\alpha f b'}{{\rm i}\omega}\Big)\,,\nn\\
&& \psi = -\fft{{\rm i}\omega\, b}{\alpha} +
\fft{f\,(r^{5-z} f b')'}{{\rm i}\omega\, \alpha r^{3(1-z)}}\,.
\eea
We can eliminate $\psi$, and obtain
\bea
\label{3matrix}
(r^{z-3-\theta} f a_1')' + \fft{\omega^2}{r^{z+5+\theta}f} a_1 &=&
\fft{Q_1}{r^{5-z+\theta}} \Big(Q_1 a_1 + Q_2 a_2-\alpha \tilde b\Big)\,,\nn\\
(r^{3z-1+\theta} f a_2')' + \fft{\omega^2}{r^{3-z-\theta}f} a_2 &=&
\fft{Q_2}{r^{5-z+\theta}} \Big(Q_1 a_1 + Q_2 a_2-\alpha \tilde b\Big)\,,\nn\\
(r^{3(z-1)} f \tilde b')' + \fft{\omega ^2}{r^{5-z} f} \tilde b &=& -\fft{\alpha}{r^{5-z+\theta}} \Big(Q_1 a_1 + Q_2 a_2-\alpha \tilde b\Big)\,,\label{3eqns}
\eea
where $\tilde b=r^{5-z} f b'/({\rm i}\omega)$.

  Note that one can again see from (\ref{3eqns}) that, as remarked previously,
it would be inconsistent to set the perturbation $a_1$ to zero, since it
would imply $\tilde b= Q_2\, a_2/\alpha$, and hence the last two equations
in (\ref{3eqns}) would be incompatible.

   For later purposes, it is useful, as in \cite{Blake:2013bqa}, to introduce
the two independent quantities
\bea
\Pi_1 &=& -r^{z-3-\theta} f a_1' -
\fft{Q_1}{\alpha} r^{3(z-1)} f \tilde b'\,,\cr
\Pi_2 &=& -r^{3z-1+\theta} f a_2' -
\fft{Q_2}{\alpha} r^{3(z-1)} f \tilde b'\,,\label{Pidefs}
\eea
which are radially conserved up to (and including)
${\cal O} (i\omega)$.\footnote{Note that the functions $\Pi_i$ are
essentially the same as the currents $\langle J_i\rangle$ in the Kubo
formula (\ref{Kubo}), since they arise as the surface terms in the
variation of the quadratic action for the fluctuations with respect to
the external sources.}
In other words, we must have
\be
\Pi_i = {\rm i} \omega\, j_i + {\cal O}(\omega^2)\,,
\qquad i=1,2.\label{piconsts}
\ee
where $j_i$ are constants.  In fact, as we shall see later in this section,
these constants are the same as the conserved currents $j_i$ introduced in
eqn (\ref{hypj1j2}).
These two conserved quantities are associated with the two zero-eigenvalue
modes of
the mass matrix
for the perturbations, which can be read off from (\ref{3matrix}).

Next, we define the two quantities
\be
H_1(\omega) = \lim_{r\rightarrow \infty}
\fft{r^{z-3-\theta} a_1'}{\, a_1}\,,\qquad
H_2(\omega) = \lim_{r\rightarrow \infty}
\fft{r^{3z-1+\theta} a_2'}{a_2}\,,\label{G1G2def}
\ee
which we associate with the following large-$r$ asymptotic behavior for $a_i$:
\bea
a_1 &=& a_{10} \Big(1 + \fft{H_1(\omega)}{(z-4-\theta) r^{z-4-\theta}} + \cdots\Big)\,,\cr
a_2 &=& a_{20} \Big(1 + \fft{H_2(\omega)}{(3z-2+\theta)r^{3z-2+\theta}} + \cdots\Big)\,.
\eea
Notice that in order for $a_1$ to be regular asymptotically, one must have
$H_{1}(\omega)=0$ when $z-4-\theta<0$.
On the other hand, the regularity of $a_2$ is guaranteed by the null
energy condition and having taken $z\geq 1$.
We will return to the vanishing of $H_1$ in more detail shortly.
In the $\omega\rightarrow 0$ limit, we can then define
\be
\gamma_i = - \lim_{\omega \rightarrow 0}
\fft{H_i(\omega)}{{\rm i}\omega}\,,\qquad i=1,2.\label{s1s2def}
\ee
The quantities $\gamma_i$ are the asymptotic data, and later, we shall
examine their relation to the conductivity matrix $\sigma_{ij}$.

   Next, let us consider the ansatz for perturbations that are purely
ingoing on the horizon, valid for small $\omega$:
\bea
a_1 &=&\fft{E_1}{\im\omega}\,
e^{-\fft{{\rm i}\omega}{4\pi T} \log f} \big(1 +
{\rm i}\omega U_1(r) + {\cal O}(\omega^2)\big)\,,\nn\\
a_2 &=& \fft{E_2}{\im\omega}\, e^{-\fft{{\rm i}\omega}{4\pi T} \log f}
\big(1 + {\rm i}\omega U_2(r) + {\cal O}(\omega^2)\big)\,,\nn\\
\tilde b& =&\fft{\nu}{\im\omega}\,
e^{-\fft{{\rm i}\omega}{4\pi T} \log f}
\big(1 + {\rm i}\omega V(r) + {\cal O}(\omega^2)\big)\,,\label{a1a2bbdef}
\eea
where the Hawking temperature $T$ is given by (\ref{hawking}).
We require $(E_1,E_2,\nu)$ to be real constants, and the $U_i$ to be
real functions that are regular both on the horizon and at asymptotic infinity.
Note that the $\im\omega$ denominators on the right-hand sides of
eqns (\ref{a1a2bbdef}) are included for convenience, in order to facilitate
the comparison with the DC ansatz approach that we described previously.
In particular, the constants $E_1$ and $E_2$ in (\ref{a1a2bbdef})
will turn out to be the same, in the $\omega\rightarrow 0$ limit, as
the constants we introduced in the DC ansatz in eqns (\ref{DCperts}).
 To see this, we recall that the
physical fluctuations of the various fields are obtained from
the complex expressions (\ref{complexrhs}) and (\ref{a1a2bbdef}) by taking the
real parts of the right-hand sides in (\ref{complexrhs}). Thus, for example,
the physical fluctuations $(\delta A_i)_x$ are given by
\bea
(\delta A_i)_x &=& \Re\Big[ \fft{E_i}{\im \omega}\,
  \Big(1-\fft{\im\omega\log f}{4\pi T} + \im\omega U_i - \im\omega t +
  {\cal O}(\omega^2)\Big)\Big] \,,\nn\\
&=& -E_i t -E_i\big(\fft{\log f}{4\pi T} - U_i\big) + {\cal O}(\omega)\,.
\eea
Taking the DC limit where $\omega\rightarrow 0$, we reproduce the
expressions for $(\delta A_i)_x$ given in (\ref{DCperts}), with
\be
\hat a_i = -E_i\big(\fft{\log f}{4\pi T} - U_i\big)\,.\label{hataiUi}
\ee
In an analogous manner we can confirm  that, as mentioned earlier,
 the constants $j_i$ appearing in
(\ref{hypj1j2}) are indeed the same as the ones arising in (\ref{piconsts}).

  Returning to the complex expressions (\ref{a1a2bbdef}), we now
substitute these into the perturbation equations (\ref{3eqns}).
    At the leading order in $\omega$, \ie at order $\omega^{-1}$, we find
\be
E_1Q_1 + E_2 Q_2 -\nu \alpha=0\,.\label{nucons}
\ee
At the next order, \ie $\omega^0$,
we have
\bea
&&(E_1 r^{z-3-\theta} f U_1')' -\fft{ Q_1}{ r^{5-z+\theta}} (E_1 Q_1 U_1 + E_2 Q_2 U_2 - \nu \alpha V) - \big(\fft{E_1 r^{z-3-\theta} f'}{4\pi T}\big)' = 0\,,\cr
&&(E_2 r^{3z-1+\theta} f U_2')' - \fft{Q_2}{r^{5-z-\theta}} (E_1 Q_1 U_1 + E_2 Q_2 U_2 - \nu \alpha V) - \big(\fft{E_2 r^{3z-1+\theta} f'}{4\pi T}\big)'=0\,,\cr
&&(\nu r^{3(z-1)} f V')' + \fft{\alpha}{r^{5-z+\theta}} (E_1 Q_1 U_1 + E_2 Q_2 U_2 - \nu \alpha V) + \big(\fft{\nu r^{3(z-1)} f'}{4\pi T}\big)'=0\,.
\eea
It follows from (\ref{piconsts}) that the first two integrals
are
\bea
\fft{r^{z-3-\theta}}{\alpha} \Big((E_1 \alpha U_1' + \nu Q_1 r^{2z+\theta} V') f -
\fft{(E_1 \alpha + \nu Q_1 r^{2z+\theta})f'}{4\pi T}\Big)&=&  -j_1\,,\cr
\fft{r^{3(z-1)}}{\alpha} \Big( (E_2 \alpha r^{2+\theta} U_2' + \nu Q_2 V') f -
\fft{(E_2 \alpha r^{2+\theta} + \nu Q_{2})f'}{4\pi T}
\Big)
&=& -j_2\,.
\eea
Evaluating the above equations on the horizon, we find
\be
j_1=\fft{\alpha E_1 + \nu\, Q_1\, r_0^{2z+\theta}}{\alpha r_0^{4+\theta}}\,,\qquad
j_2=\fft{1}{\alpha}(\nu\, Q_2 + \alpha\, E_2\, r_0^{2+\theta}) \,
r_0^{2z-4}\,.\label{c1c2value}
\ee
Using (\ref{nucons}) to substitute for $\nu$ in these equations, we
obtain expressions for the $j_i$ in terms of the $E_i$ which are precisely
those given by eqns  (\ref{condmatrix}) and (\ref{sigcomps}).

   The calculation above shows how the conductivities are read off from
the horizon data.  We next turn to a discussion of how they are related to
data on the boundary at infinity.  In particular, we shall see that
regularity requirements at the boundary can provide additional
constraints on the currents $j_1$ and $j_2$, and hence modify the
conductivity matrix.

   To calculate the quantities $\gamma_i$ defined in (\ref{s1s2def})
and (\ref{G1G2def}),
we first take the $\omega=0$ limit, and define the functions $W_1$ and $W_2$
by
\bea
\lim_{\omega\rightarrow 0} \fft{r^{z-3-\theta} a_1'}{(-{\rm i}\omega\, a_1)}
&=&
r^{z-3-\theta}\Big(\fft{f'}{4\pi T\,f} - U_1'\Big)\equiv r^{2(z-1)} W_1\,,\cr
\lim_{\omega\rightarrow 0} \fft{r^{3z-1+\theta} a_2'}{(-{\rm i}\omega\, a_2)}
&=& r^{3z-1+\theta}\Big(\fft{f'}{4\pi T\,f} - U_2'\Big)\equiv r^{2(z-1)}
W_2\,.\label{w1w2def}
\eea
It turns out that $W_1$ and $W_2$ satisfy
\bea
(r^{3z+1+\theta} f^2\, W_1')' &=&\fft{
j_1 \,(Q_2^2 + \alpha^2 \,r^{2+\theta})-
j_2\, Q_1\, Q_2}{E_1 \,r^{z+1+2\theta}}\,,\cr
(r^{3z+1+\theta} f^2 \,W_2')' &=&-\fft{j_2\,\alpha^2 +
Q_1 \,(j_2 \,Q_1 - j_1\, Q_2) \,r^{2z+\theta}}{E_2 \,r^{z-1}}\,,
\label{w1w2eom2nd}
\eea
which in turn imply that
\bea
r^{3z+1+\theta} f^2 \,W_1' &=& d_1 - \fft{1}{E_1\, r^{z+\theta}}\Big(
\fft{Q_2\,(j_1\,Q_2 - j_2 \,Q_1)}{z+\theta} +
\fft{\alpha^2 \,j_1 \,r^{2+\theta}}{z-2}\Big) \equiv\xi_1\,,\cr
r^{3z+1+\theta} f^2 \,W_2' &=& d_2 -
\fft{r^2}{E_2}\Big(\fft{Q_1\, (j_2\,Q_1-j_1\, Q_2) \,r^{z+\theta}}{z+2+\theta} -
\fft{\alpha^2 \,j_2}{(z-2) \,r^z}\Big) \equiv \xi_2\,,
\eea
where $d_1$ and $d_2$ are integration constants.
Together with (\ref{w1w2def}), we have
\bea
(\fft{U_1'}{r^{z+1+\theta}})' &=& \tilde\zeta_1 \equiv \Big(\fft{f'}{4\pi T\, r^{z+1+\theta} f}\Big)' - \fft{\xi_1}{r^{3z+1+\theta}f^2}\,,\cr
(r^{z+1+\theta} U_2')' &=& \tilde\zeta_2 \equiv   \Big(\fft{r^{z+1+\theta} f'}{4\pi T\, f}\Big)' - \fft{\xi_2}{r^{3z+1+\theta}f^2}\,.
\eea
It turns out that by choosing the integration constants appropriately,
the singularity at $r=r_0$ in the function $\tilde\zeta_i$ can be avoided.
This ensures that $U_1$ and $U_2$ are regular on the horizon.
The leading-order
large-$r$ expansions for $\tilde\zeta_1$ and $\tilde\zeta_2$ depend upon the
interval in which the Lifshitz exponent $z$ lies.  We find
\bea
\tilde\zeta_1 &=&\left\{
          \begin{array}{ll}
            -\fft{j_1\alpha^2}{(z-2)E_1} \big(\fft{1}{r}\big)^{4z-1+\theta} + \cdots, &\qquad 1\le z<2;  \\
            -\fft{d_1}{r^{3z+1+\theta}} + \cdots, &\qquad 2<z\le 4; \\
            \fft{\rm const.}{r^{2z+5+\theta}} + \cdots, &\qquad z > 4;
          \end{array}
        \right.\\
\tilde\zeta_2 &=&
\left\{
  \begin{array}{ll}
    -\fft{2(z-1) (E_1 Q_2 - E_2 Q_1 r_0^{2z+2+2\theta})}{E_2 Q_1 r_0^{4+\theta}} \fft{1}{r^{2z-1}} + \cdots, &\qquad z<2 \\
    \fft{\rm const.}{r^3} + \cdots, &\qquad z>2 \\
  \end{array}
\right.
\eea
which imply that for $z>1$ and $\theta>0$,  $\tilde\zeta_i$ can be integrated out
to infinity without divergence.
Thus the general solutions for $U_i'$ are  given by
\bea
U_1' = r^{z+1+\theta} \big(\beta_1 + \int_\infty^r \tilde\zeta_1\big)\,,\qquad
U_2' = \fft{1}{r^{z+1+\theta}} \big(\beta_2 + \int_\infty^r \tilde\zeta_2\big)\,,
\eea
where the $\beta_i$'s are two integration constants.
It is clear that the regularity of $U_1$ at asymptotic infinity requires that $\beta_1=0$.

We are now in a position to obtain the two quantities
\bea
\gamma_1 &=& \lim_{\omega\rightarrow 0}^{r\rightarrow \infty}
\fft{r^{z-3-\theta} a_1'}{(-{\rm i}\omega\, a_1)}=0\,,\nn\\
\gamma_2 &=& \lim_{\omega\rightarrow 0}^{r\rightarrow \infty}
\fft{r^{3z-1+\theta} a_2'}{(-{\rm i}\omega\, a_{2})}
=\fft{j_2\, Q_1-j_1\, Q_2}{E_2 Q_1}=
r_0^{(2z-2+\theta)} \Big(1 - \fft{E_1Q_2}{E_2Q_1} \fft{1}{r_0^{2(z+1+\theta)}}\Big)\,,\label{res1}
\eea
working under the assumption that
$z>\ft43$.
Indeed, for $1\le z \le \ft43$, we find that the the leading falloff of $U_1$
becomes divergent.  Specifically, for $1\le z<\ft43$ we find
\be
U_1 = -\fft{\alpha^2 j_1}{(z-2)(4z-2+\theta) (3z-4)} r^{4-3z} + \cdots\,,
\ee
whilst for $z=\ft43$ we find
\be
U_1 = -\fft{9 \alpha^2\, j_1}{2(10+3\theta)}\, \log r+\cdots\,.
\ee
The convergence of $U_1$ at large $r$ for $1\le z\le \ft 43$ requires
either $\alpha=0$ or $j_1=0$.
Since we are interested in the effect of a nonzero $\alpha$, for now we consider $j_1=0$.
It follows from (\ref{c1c2value}) that
\be
\gamma_2 = r_0^{2z-2+\theta} \Big(1 + \fft{Q_2^2}{r_0^{2+\theta}
(\alpha^2 + Q_1^2 \, r_0^{2z+\theta})}\Big)\,,
\qquad  1 \le z \le \fft43\,.\label{res2}
\ee
Note that in the AdS limit where $z=1$ and $\theta=0$, and hence $Q_1=0$, we successfully reproduce the previous known result in the literature.

On the other hand if we have $\alpha=0$, then we can have all $z\ge 1$, including $z=2$.
It follows from (\ref{nucons}) and (\ref{res1}) that
\be
\gamma_2 = r_0^{2z-2+\theta}
\Big(1 - \fft{Q_2^2}{Q_1^2\,  r_0^{2z+2+\theta}}\Big)\,.\label{alp=0sigdc2}
\ee
This result is applicable for all $z\ge 1$.  It coincides
with (\ref{res2}) when $1 \le z\le 4/3$.

    It is interesting to examine how the two asymptotically-defined
quantities
$\gamma_{i}$ and $\psi_\infty$ are
related to the currents $j_i$. It follows from
(\ref{G1G2def}), (\ref{s1s2def}) and (\ref{w1w2def}) that
\be
\gamma_1 = \lim_{r\rightarrow \infty} r^{z-3-\theta}\Big(\fft{f'}{4\pi T\,f} - U_1'\Big)\,,\qquad
\gamma_2 = \lim_{r\rightarrow \infty}r^{3z-1+\theta}
\Big(\fft{f'}{4\pi T\,f} - U_2'\Big)\,.
\ee
We can then use (\ref{hataiUi}) to obtain
\be
\gamma_1= -\lim_{r\rightarrow \infty} r^{z-3-\theta}
\fft{\hat a_1'}{E_1}\,,\qquad
\gamma_2 = -\lim_{r\rightarrow \infty}r^{3z-1+\theta}
\fft{\hat a_2'}{E_2}\,.
\ee
It now follows from (\ref{hypj1j2}) that
\be
\gamma_1= \lim_{r\rightarrow \infty}
  \fft{j_1 + Q_1 \,\hat \psi(r) }{E_1 f(r)}\,,\qquad
\gamma_2 = \lim_{r\rightarrow \infty}
\fft{j_2 + Q_2\, \hat \psi(r) }{E_2 f(r)}\,.
\ee
Since $f(\infty)=1$, we find the following relation between the boundary
quantities $\gamma_i$ and the conserved currents:
\be
\gamma_1=\fft{j_1}{E_1} + \fft{Q_1\,\hat \psi_\infty}{E_1}\,,\qquad
\gamma_2=\fft{j_2}{E_2} + \fft{Q_2\,\hat \psi_\infty}{E_2}\,.
\ee
Recalling that $\hat \psi_\infty=-\frac{j_1}{Q_1}$, it is now clear that
$\gamma_1 = 0$, regardless of whether $j_1=0$ or not.
Moreover, when $j_1=0$ we recover the result $\gamma_2 = j_2/E_2$,
from which we can immediately conclude that
in this case $\gamma_2$ is precisely the one we found in eqn
(\ref{jzerosigmaDC}).

\section{Temperature Dependence}
\label{Tdep}

We are now ready to return to the issue of the temperature dependence
of the conductivities.
In terms of the horizon radius $r_0$ and the parameters $\alpha$ and $Q_2$, the Hawking temperature (\ref{hawking})
is given by
\be
\label{fullT}
T=\fft{z+2 + \theta}{4\pi} r_0^z - \fft{Q_2^2}{8\pi (2+\theta)}\fft{1}{r_0^{z+2+2\theta}} -
\fft{\alpha^2}{4\pi(2 + \theta)} \fft{1}{r_0^{z+\theta}}\, .
\ee
When $r_0^{2z+\theta} >> \alpha^2 $ and $r_0^{2z+2+2\theta} >> Q_2^2$, we recover
the well known Lifshitz scaling\footnote{This is also the correct relation when $\alpha=Q_2=0$.}
\be
\label{TLif}
T\sim r_0^z\,.
\ee
Thus, this approximation corresponds to ``large temperatures,'' in the sense of \footnote{Note that the temperatures satisfying
these ranges can be decreased/increased by tuning $\alpha$ and $Q_2$.}
\be
\label{largeTapprox}
T>> \alpha^{\frac{2z}{2z+\theta}}  \quad \text{and} \quad T>>Q_2^{\frac{z}{z+1+\theta}} \, .
\ee
As the temperature is lowered, at some point the $\alpha^2$ term in
(\ref{fullT})  (which always dominates over the $Q_2^2$ term provided that
$\theta>-2$)
will have to be taken into account. At that point, the expression for
temperature in terms of $r_0$ takes the form
\be
\label{subleadingT}
T \sim \fft{z+2 + \theta}{4\pi} r_0^z -
\fft{\alpha^2}{4\pi(2 + \theta)} \fft{1}{r_0^{z+\theta}}\,,
\ee
which holds under the condition $ \alpha^2 \, T^{\frac{2+\theta}{z}} >> Q_2^2$.
Note that the relation (\ref{subleadingT}) becomes \emph{exact} when $Q_2=0$.

In the large temperature limit -- or alternatively, in the small $\alpha$ limit -- we can invert (\ref{subleadingT}) to obtain an expansion for
$r_0$ as a function of $T$,
\be
\label{correctedr0}
r_0 = \Big(\fft{4\pi T}{z+2+\theta}\Big)^{\fft1{z}} +
   \fft{\alpha^2}{4\pi z (2+\theta)}\, \fft{1}{T}\,
   \Big(\fft{z+2+\theta}{4\pi T}\Big)^{1+\fft{\theta-1}{z}}+\cdots\,.
\ee
For example, if $z=\theta=1$ we have
\be
r_0 = \pi \, T + \fft{\alpha^2}{12\pi^2\,  T^2} +\cdots\,.
\ee
We shall return to the opposite, low temperature, regime shortly.
Using the large $T$ approximation (\ref{correctedr0}), the conductivity matrix $\sigma_{ij}$ given in (\ref{sigcomps}) becomes
\bea
\sigma_{11}&\sim& \frac{Q_1^2}{\alpha^2} \, T^{\fft{2z-4}{z}}+
T^{-\frac{(4+\theta)}{z}} + \ldots \;,\cr
\sigma_{12}&\sim& Q_1Q_2 \left(\fft{1} {\alpha^2} \, T^{\fft{2z-4}{z}} +
(z-2) \, T^{-\frac{(4+\theta)}{z}} \right) + \ldots \;, \cr
\sigma_{22} &\sim& T^{\fft{2z-2+\theta}{z}} +
\frac{Q_2^2}{\alpha^2} \, T^{\frac{2z-4}{z}} +
Q_2^2 \, (z-2) \, T^{-\frac{(4+\theta)}{z}} + \ldots \; ,
\eea
where each term in each component is smaller than the preceding one. Note that we
are dropping all the strictly positive numerical factors that don't involve the parameters $\{Q_1,Q_2, \alpha\}$.
The terms containing $\alpha$ are the leading effects due to the momentum relaxation mechanism triggered by the axions, and
in the large temperature limit (\ref{TLif}) they all scale like
$\sim T^{\frac{2z-4}{z}}$.
Note that when $z=4/3$ these terms yield contributions proportional to $\sim T^{-1}$,
which in turn can lead to a resistivity linear in temperature, as we will see shortly.
In order to understand the temperature behavior over a wider range one must invert (\ref{fullT}) for generic values of the scaling exponents.
While this can be done numerically, it is beyond the scope of this paper.
Finally, we note that while the $Q_1$ charge naively sets a scale that is different from those controlling the temperature, which are $Q_2$ and $\alpha$,
it is fully determined by the background, \emph{i.e.} it is fixed in
terms of $z$ and $\theta$, as given in (\ref{parameters}). .

Recall that when $Q_2=0$ the DC conductivity in the system is simply $\sigma_{11}$ given above.
Written in terms of the conduction exponent $\zeta_1=-2-\theta$
defined in (\ref{zeta1}) and the parameters (\ref{parameters}) describing our solution,
the temperature dependence associated with $\sigma_{11}$ is
\be
\sigma_1^{\rm DC} \sim T^{\frac{\zeta_1-2}{z}} + \frac{Q_1^2}{\alpha^2} \, T^{\fft{\zeta_1-\lambda_3 \gamma}{z}} + \ldots \;
= \; T^{\frac{\zeta_1-2}{z}} \left( 1 + \frac{Q_1^2}{\alpha^2} \, T^{\frac{2z+\theta}{z}}\right) + \ldots
\; ,\label{zetaform}
\ee
and matches the generic behavior observed\footnote{More specifically,
when $z>1$ our expression for
$\sigma_{11}$ falls into Class III of \cite{Gouteraux:2014hca}.
Note that our $\theta$ is opposite to that of
\cite{Gouteraux:2014hca}, since our radial coordinate is the inverse
of the one used there.} in \cite{Gouteraux:2014hca}.
Let us now inspect more carefully the structure of (\ref{zetaform}),
and ask in particular whether it allows for the scaling
$\sigma \sim 1/T$ so that the associated resistivity  is of the form $\rho \sim T$.
First, note that the $\sim Q^2_1/\alpha^2$ term in (\ref{zetaform}) will always dominate over the first one
when (\ref{largeTapprox}) holds, since $Q_1 \sim {\cal O}(1)$.
Thus, even though $T^{\frac{\zeta_1-2}{z}}\sim T^{-1}$
when $z=4+\theta$, this scaling is subleading
and therefore not visible.
This conclusion would change if we had $\theta<0$, which for our solutions was
 forbidden in order to avoid
UV curvature singularities (note for example that when $\theta = -2z$ both
terms in $\sigma_{1}^{\rm DC}$ have the same temperature dependence, for all $T$).
Indeed, one should keep in mind that the restriction $\theta>0$ would be relaxed
if the scaling solutions we are studying were
embedded in AdS -- thus changing the UV of the theory but leaving the horizon $\{z,\theta\}$ scaling behavior untouched.
Such constructions -- which entail appropriately modifying the scalar potential to allow the scalar field $\phi$ to settle
to a constant at the boundary --
would then admit a wider range for $\theta$, including negative values.
For appropriate choices of parameters, these kinds of solutions can be constructed numerically. However, we won't attempt to do so here.

The dominant, $\alpha$-dependent contribution
to the conductivity does scale as $1/T$ when we choose $z=4/3$.
Indeed, this case yields
\be
\label{sigma1largeT}
 \sigma_1^{\rm DC}  \sim \frac{Q_1^2}{\alpha^2} \, \frac{1}{T} + \frac{1}{T^{3+3\theta/4}}
\quad \Rightarrow \quad
\rho \sim \frac{T^{3+\frac{3\theta}{4}}}{1+\frac{Q_1^2}{\alpha^2}T^{2+\frac{3\theta}{4}}} \; ,
\ee
and is therefore  associated -- at leading order -- with a linear resistivity $\rho\sim T$.
Interestingly, the $z=4/3$ case
is precisely the situation for which the gauge field perturbation becomes singular, as discussed in Section 4.
Thus, to describe a resistivity that is (nearly) linear in temperature we should take $z = 4/3 - 4\epsilon/9$,
with $\epsilon <<1$, for which we would then have
\be
\sigma \sim \frac{1}{T^{1+\epsilon}}  + \ldots \; .
\ee
Alternatively, embedding these solutions in AdS, as described above, would modify the asymptotic structure of the perturbations, and allow
one to take exactly $z=4/3$.

As the temperature becomes smaller, the (nearly) linear behavior is modified by the presence of the
second term in the conductivity (\ref{sigma1largeT}), which starts to dominate.
To determine precisely its fall off (which is controlled by $\theta$)
one must go beyond the large temperature approximation.
On the other hand, we stress once again that the situation would be different if negative values of $\theta$ were allowed, as would be the case
if these solutions were embedded in AdS.
A particularly interesting case is that of $\theta = -2z$, for which the relation (\ref{fullT}) yields the exact expression
\be
\label{exactT}
T=\left[ \fft{z+2 + \theta}{4\pi} - \fft{\alpha^2}{4\pi(2 + \theta)} \right] r_0^z \, ,
\ee
and so the temperature scaling is precisely $T\sim r_0^z$.
When $z=4/3$ (and thus $\theta=-8/3$) it can be easily shown that both terms in $\sigma_1^{\rm DC}$
scale as $1/T$, and therefore the resistivity is always linear.\footnote{Note that in this case, we have
$f(r\rightarrow \infty) \sim Q_2 r^{\fft23}$.  This is however not a problem in this case since the solution is
viewed as an approximation in the IR region,
expected to be embedded in the asymptotic AdS spacetimes.}

Another interesting choice is that of $z=4/3$ and $\theta=-z$, which yields
\be
\label{sigma1linquad}
\sigma_1^{\rm DC} \sim \frac{\alpha^2 + Q_1^2 T}{\alpha^2 T^2} \, ,
\ee
and therefore a quadratic regime for the resistivity below the linear one.
Intriguingly, the case $z=4/3,\ \theta=-4/3$ is partially reminiscent of that
singled out by \cite{Hartnoll:2015sea}, whose analysis relied on
purely field-theoretic arguments.
A comparison between their setup and ours leads to the identification $\Phi = \zeta_1-\theta-2$ for the
$\Phi$ parameter of \cite{Hartnoll:2015sea}.
Since for our solutions the conduction exponent is not free, but is constrained to be $\zeta_1 = -2-\theta$,
we obtain the simple identification $\Phi =\zeta_1$.
Thus, the special value $\Phi=-2/3$ singled out by \cite{Hartnoll:2015sea} corresponds for our background
to the $\theta = -4/3$ case leading to (\ref{sigma1linquad}).
However, recall that in \cite{Hartnoll:2015sea} the authors had obtained $\theta=0$, which is not what we have.
For our particular single-charge solutions in the large temperature limit,
the choice $\Phi = -2/3,\ z=4/3$ with $\theta=0$ is not allowed.
Still, we wonder whether the fact that the value $z=4/3$ plays a crucial role in our solutions is more than a mere coincidence.

That the choices $\{ z=4/3,\ \theta=-8/3\}$ as well as
$\{z=4/3,\ \theta=-4/3\}$
are allowed by all energy and stability conditions, provided one has AdS asymptotics, can be seen from the
parameter ranges summarized \emph{e.g.} in \cite{Cremonini:2016bqw}.
As we already noted, in full generality one should solve for the temperature dependence numerically, since subleading
effects will become crucial when going to even smaller temperatures.

We now examine the structure of $\sigma_2^{\rm DC}$ given by (\ref{jzerosigmaDC}).
Working again in the large $T$ approximation (\ref{TLif}), we see that
\be
\label{sigma2Temp}
\sigma_2^{\rm DC} \sim T^{\fft{2z-2+\theta}{z}} \left[1 +
\fft{Q_2^2}{T^{\fft{2+\theta}{z}} \left(\alpha^2 +
Q_1^2\, T^{\fft{2z+\theta}{z}}\right)}\right] \, .
\ee
Since in the range (\ref{largeTapprox}) the contribution from the $\alpha$ term in the denominator is always subleading compared to that of $Q_1$
 -- recall that the charge $Q_1$ is fixed by the background and is ${\cal O}(1)$ -- the leading terms in the expansion of $\sigma_2^{\rm DC}$
 are given by
\be
\label{sigma2zeta}
\sigma_2^{\rm DC} \sim  T^{\fft{2z-2+\theta}{z}} +  \frac{Q_2^2}{Q_1^2} \, T^{\frac{-\theta-4}{z}} + \ldots  =
T^{\frac{\zeta_2-2}{z}} + \frac{Q_2^2}{Q_1^2} \, T^{\frac{\zeta_1-2}{z}}  +
\ldots \; ,
\ee
and are therefore insensitive to the magnitude $\alpha$ of the axionic sources.
Notice that a linear regime could arise from the term $\propto Q_2$ when $z=\theta+4$.
However, it would occur at best in a narrow temperature range, since the first term
dominates the large $T$ behavior, as its power is constrained to be positive by the null energy conditions.
The contribution encoded by $\alpha$ will then come into play when we consider subleading terms.
As an example, we can examine the form of $\sigma_2^{\rm DC}$ in the small $\alpha$ approximation (\ref{subleadingT}).
Starting from the expression (\ref{jzerosigmaDC}) and expanding to linear order in $\alpha^2$, we find
\be
\sigma_2^{\rm DC} \sim T^{\fft{2z-2+\theta}{z}} + \frac{Q_2^2}{Q_1^2} \, T^{\frac{-\theta-4}{z}}
+ \alpha^2 T^{-\frac{2}{z}} - \frac{\alpha^2 Q_2^2}{Q_1^2} T^{-\frac{2}{z}(2+z+\theta)} \ldots \; ,
\ee
where we are suppressing positive coefficients that depend on $(z,\theta)$.
The competition between the different terms in the expansion will then be sensitive to the size of $Q_2$ and $\alpha$ as well as the
particular values of the scaling exponents.
Whether $\sigma_2^{\rm DC}$ can give rise to a linear DC resistivity in other regimes entails a detailed study of the relation (\ref{fullT}).

Inspecting the behavior of $\sigma_2^{\rm DC}$ for generic temperatures, we see that it
differs in a crucial way from the result of \cite{Ge:2016lyn}.
The $Q_1^2$ term in the denominator of our expression (\ref{sigma2Temp}),
which is temperature dependent, does not appear in \cite{Ge:2016lyn}, precisely because both gauge
fields were not allowed to fluctuate in their analysis.
One cannot merely set $Q_1=0$ in this expression, without also setting $z=1$.
Furthermore, by naively suppressing the $Q_1$ term and incorrectly identifying $\sigma_2^{\rm DC}$ with
\be
\sigma_{22} \sim  T^{\fft{2z-2+\theta}{z}} +  \frac{Q_2^2}{\alpha^2} \, T^{\frac{2z-4}{z}} + \ldots \; ,
\ee
one is in fact turning off a contribution that is more important
-- in the large temperature regime (\ref{TLif}) used to arrive at this expression -- than that
coming from the $\alpha$ term.
Moreover, notice that the $z=1$ and $\theta=-1$ case considered in \cite{Ge:2016lyn}, which naively yields
a linear temperature dependence for
the resistivity, $\sigma_{22}(z=1,\ \theta=-1)
\sim \frac{1}{T} + \frac{Q_2^2}{\alpha^2} \frac{1}{T^2}$,
violates the null energy conditions and is therefore problematic.

Finally, we examine numerically the temperature dependence of $\sigma_{2}^{\rm DC}$ for specific scaling exponents,
using the exact expressions (\ref{jzerosigmaDC}) and (\ref{fullT}),
to get a better feel for some of the possible behaviors allowed by our model.
We focus on the case
where the hyperscaling-violating solutions are treated as the full
solutions from the horizon to asymptotic infinity, rather than as
approximate solutions in the IR region.  As we have discussed earlier,
this implies that $j_1=0$ for $1\le z\le \fft43$.  The
$\sigma_2^{\rm DC}$ conductivity is then given by (\ref{sigma2Temp}).
As a concrete example, we consider the case $z=\fft43$ and $\theta=0$, and choose the
parameters $Q_2=1000$ and $\alpha=10$.  
For $T\le T_0\sim 17.5$, the resistivity can be approximated by a linear
function with
\be
\rho_2=0.0222 (1 + 0.283 T)\,,\label{rho2Tlinear1} \,
\ee
and at $T=T_0$ it turns around\footnote{Similar transitions were seen in massive gravity in \cite{Baggioli:2014roa}.},
\be
\fft{\partial \rho_2}{\partial T}\Big|_{T=T_0}=0\,.
\ee
For $T>T_0$ it is convenient to approximate it using the Steinhart-Hart equation
\cite{steinhart},
\be
\fft{1}{T} = A + B \log(\rho_2) + C (\log(\rho_2))^3 +
D (\log(\rho_2))^2\,,\label{steinhart}
\ee
with 
coefficients
\be
A=0.818\,,\qquad B=0.702\,,\qquad C=0.0202\,,\qquad D=0.205\,.
\ee
We plot $\rho_2(T)$ and the two approximate functions in Fig.~1.

\begin{figure}[ht!]
\begin{center}
\includegraphics[width=300pt]{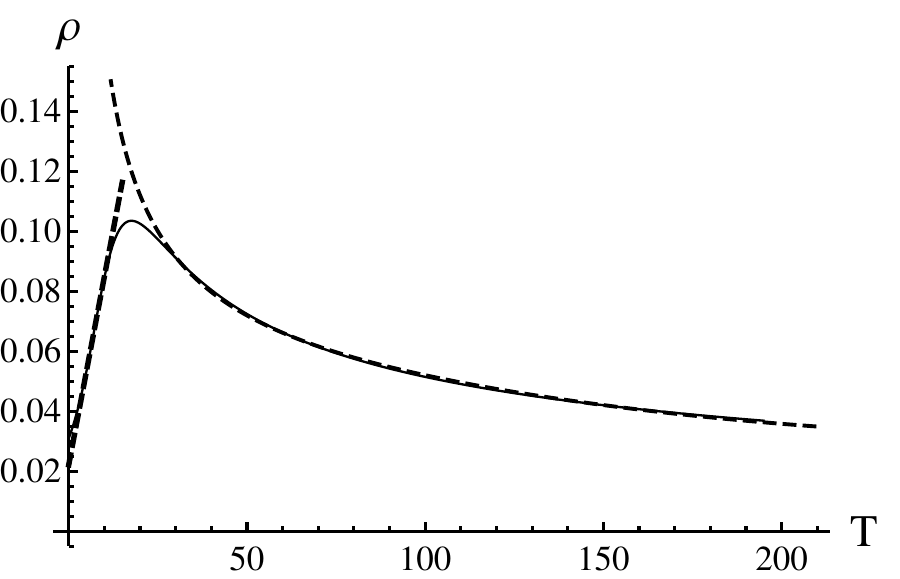}\ \
\end{center}
\caption{\it Plot of resistivity versus temperature for $z=4/3,\theta=0, Q_2=1000$ and $\alpha=10$, showing a
transition from a linear regime to a decaying region. The resistivity $\rho_2(T)$ is plotted as a solid thin line.
The straight dashed line on the left represents
(\ref{rho2Tlinear1}), while the curved dashed line on the right represents
(\ref{steinhart}).}
\end{figure}

However, for this choice of parameters the validity of the linear approximation at small $T$ is rather restricted.  
As we can see in Fig.~2 where the low temperature region is magnified, the straight line (\ref{rho2Tlinear1}) 
appearing in Fig.~1 matches with the resistivity roughly between $T\in (4.0,10)$.  
As smaller temperatures we have instead 
$\rho_2 \sim 0.03 (1 + 0.1 T + 0.01 T^2 + 0.00006 T^3)$ and the third-order term can be ignored.
In the higher-temperature phase the
approximation to (\ref{steinhart}) is accurate over a wide range.  
Note also that in this example we have chosen $Q_2=1000$ and $\alpha=10$, with a large ratio $Q_2/\alpha=100$.  
If we consider a smaller value $Q_2=10$, so that $Q_2/\alpha=1$ instead, the nearly linear small $T$ region disappears and once finds 
the decaying behavior seen in Fig.~3. 
Note that for the case with $\alpha=0$, corresponding to an infinite ratio of $Q_2/\alpha$, we get the $\rho_2$ temperature 
dependence shown in Fig.~1.   
These features appear somewhat robust to changes in $z$ and $\theta$.

\begin{figure}[ht!]
\begin{center}
\includegraphics[width=300pt]{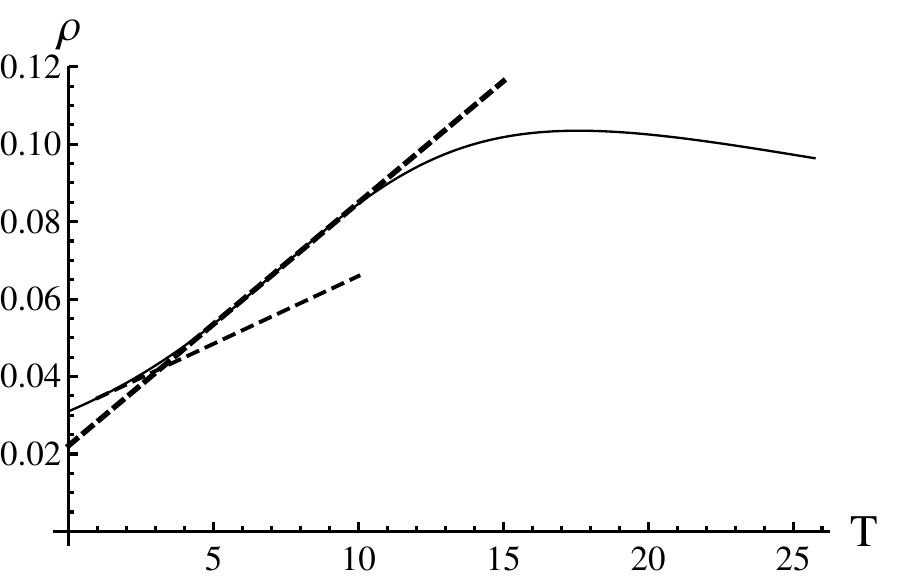}\ \
\end{center}
\caption{\it Plot of resistivity versus low temperature for $z=4/3,\theta=0, Q_2=1000$ and $\alpha=10$, The longer straight dashed line represents
(\ref{rho2Tlinear1}).}
\end{figure}

\begin{figure}[ht!]
\begin{center}
\includegraphics[width=300pt]{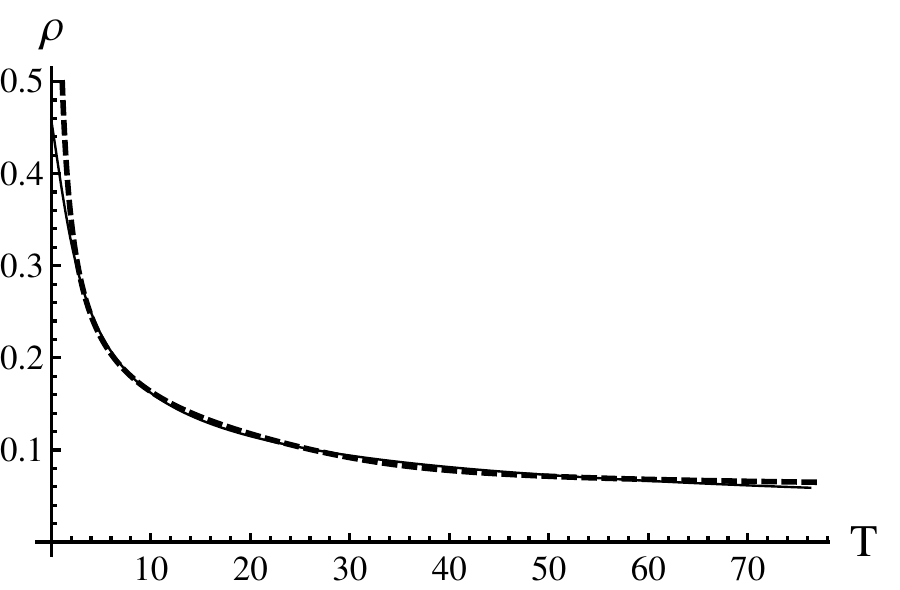}\ \
\end{center}
\caption{\it Plot of resistivity versus low temperature for $z=4/3,\theta=0, Q_2=10$ and $\alpha=10$. 
The dashed curved line represents (\ref{steinhart}) with Steinhart-Hart coefficients $A=2.29, B=2.67, C=0.145, D=1.07$. }
\end{figure}

We close this section with a few comments on the extremal limit for which $T=0$.
Let us denote the horizon radius by $r_0=\bar r_0$.  We have (for $T\sim 0$) that
\be
\fft{z+2 + \theta}{4\pi} \bar r_0^z=  \fft{Q_2^2}{8\pi (2+\theta)\,
\bar r_0^{z+2+2\theta}} +
\fft{\alpha^2}{4\pi(2 + \theta)\, \bar r_0^{z+\theta}}\,.
\ee
We can solve for $Q_2$ in terms of $\bar r_0$ and $\alpha^2$.  It is then
easy to establish that for small $T$, we have
\be
r_0=\bar r_0 + \fft{4\pi \bar r_0^{z+1+\theta}}{[-\alpha^2 +
2(z+1+\theta)(z+2+\theta) \, \bar r_0^{2z+\theta}]}\, T\,.
\ee
Thus, perhaps not surprisingly, we have
\be
\sigma_{ij} \sim \bar\sigma_{ij} + {\cal O}(T)\,.
\ee
This linear dependence can be seen as the shorter dashed line in Fig.~2.

\section{Conclusions and Summary of Results}

In an attempt to gain insight into strongly coupled phases with anomalous scalings, we have chosen to work with an
holographic model that gives rise to non-relativistic geometries that violate hyperscaling.
These provide a fruitful laboratory for realizing geometrically a variety of scalings, and insights into the potential mechanisms behind them.
The solutions we have examined are supported by a running dilatonic scalar and two gauge fields, with the latter playing very different roles.
One gauge field is responsible for generating the Lifshitz-like nature of the background, with its charge $Q_1$ entirely fixed in terms of the scaling
exponents.
The other one plays a role analogous to that of a standard Maxwell field in asymptotically AdS space,
and its charge $Q_2$ is a free parameter.
Since our interest here is in the computation of DC conductivities, we have
included two spatially dependent axionic fields which encode the physics of momentum dissipation in the dual theory,
without spoiling the homogeneity of the background.
Consistency of the resulting perturbation equations requires both gauge
fields to fluctuate, which leads to some subtleties in the analysis.

As we have seen, the conductive response of the system to turning on two electric fields is characterized by a matrix of conductivities $\sigma_{ij}$
whose components are
\be
\sigma_{11} =\fft{1}{r_0^{4+\theta}} + \fft{Q_1^2 \,r_0^{2z-4}}{\alpha^2}\,, \;\;
\sigma_{12} =\sigma_{21}=\fft{Q_1Q_2r_0^{2z-4}}{\alpha^2}\,, \;\;
\sigma_{22}=r_0^{2z-2+\theta} + \fft{Q_2^2}{\alpha^2} r_0^{2z-4}\,,
\ee
with $\alpha$ the magnitude of the axionic scalars.
The temperature dependence is therefore controlled by the interplay between the horizon size and the three quantities $Q_1$ (fixed by the background),
$Q_2$ and $\alpha$.
In particular, in the ``large temperature'' regime $r_0^{2z+\theta} >> \alpha^2$ and $r_0^{2z+2+2\theta} >> Q_2^2$
we recover the standard Lifshitz scaling $T\sim r_0^z$, while subleading temperature effects are encoded by (\ref{subleadingT}).
A detailed study of the temperature dependence can be performed numerically by inverting expression (\ref{fullT}) once the scaling exponents are specified,
but is not feasible analytically in full generality.

The main differences between our setup, in which the solutions are Lifshitz-like even in the UV, and the more standard case with AdS asymptotics
comes from examining the boundary behavior of the perturbations.
Ensuring a well-behaved boundary expansion
requires taking the current associated with the Lifshitz gauge field to vanish when $1 \leq z \leq 4/3$.
Moreover, avoiding UV curvature singularities requires $\theta$ to be positive. These constraints on the scaling exponents
can be relaxed, however, by embedding the solutions in AdS.
While this can be done numerically for particular choices of parameters, one loses the clear advantage of working with somewhat simple analytical
solutions.
Still, this is an important point to keep in mind, since an AdS embedding modifies the UV properties without affecting the horizon behavior.
In this paper we haven't attempted to realize such geometries in AdS, but have focused instead on examining the relation between
horizon and boundary data for Lifshitz asymptotics, and the role of the two distinct gauge fields.

In the simple single charge case $Q_2=0$,
the DC conductivity in the large temperature regime described above takes the form
\be
\sigma_1^{\rm DC} = \sigma_{11} \sim \frac{Q_1^2}{\alpha^2} \, T^{\frac{2z-4}{z}} + T^{-\frac{4+\theta}{z}} \, ,
\label{sigma1conclusions}
\ee
which we note is in agreement with the IR analysis of \cite{Gouteraux:2014hca}.
For the special value $z=4/3$, the leading term gives rise to a
linear resistivity regime $$\rho \sim T \, , $$
with the subleading behavior controlled by $\theta$.
While the precise value $z=4/3$ is strictly forbidden for Lifshitz
asymptotics (but is allowed in AdS embeddings), a \emph{nearly linear} regime can be achieved
by taking the dynamical critical exponent to be arbitrarily close to it.
In addition, the restriction $\theta>0$ that comes about from avoiding UV curvature singularities
is relaxed by AdS boundary conditions, changing the resulting phenomenology at smaller temperatures, as discussed in Section 5.

An important point to keep in mind is that the
range of temperatures for which the result (\ref{sigma1conclusions}) and its two-charge generalizations apply can be tuned by
adjusting the two parameters $\alpha$ and $Q_2$ as desired. Indeed, ``large'' temperatures are only large compared to
appropriate powers of $\alpha$ and $Q_2$, and therefore one can push the linear resistivity regime to smaller or larger temperatures
by simply changing the size of these two tunable parameters. The existence of the additional scale set by $Q_2$
is one of the advantages of working with a model that involves two gauge fields.
An additional feature to emphasize is that in our model the $\{z,\theta\}$ scaling solutions occupy the entire geometry and not just
its IR portion. As a result, they can in principle describe intermediate scalings (much as in \cite{Bhattacharya:2014dea}, where the focus however
was on the behavior of the optical conductivity).

When the current associated with the gauge field
that generates the Lifshitz background vanishes (which is required
when $1\leq z \leq 4/3$),
the DC conductivity in the temperature regime $T\sim r_0^z$
is given by
\be
\label{sigma2Temp2}
\sigma_2^{\rm DC} \sim T^{\fft{2z-2+\theta}{z}} \left[1 +
\fft{Q_2^2}{T^{\fft{2+\theta}{z}} \left(\alpha^2 +
Q_1^2\, T^{\fft{2z+\theta}{z}}\right)}\right] \, ,
\ee
and is controlled by the interplay between the two charges and the size of the axionic fields.
Unlike in the single charge case discussed above, we find that when $\theta >0$ the leading order, large temperature behavior
does not allow for a linear resistivity (it does however come about by allowing $\theta<0$).
Whether it can be generated for other temperature regimes is an interesting question, and
indeed we see evidence that $\rho \sim T$ arises at lower temperatures, as shown in Fig.~1.
As an example, we have studied numerically the temperature dependence of $\sigma_2^{\rm DC}$ for $z=4/3$ and $\theta=0$,
using the exact expressions (\ref{jzerosigmaDC}) and (\ref{fullT}).
For sufficiently large $Q_2/\alpha$ ratio, we have seen that
the system describes a transition from a linear regime at low temperature, to a decaying regime at higher temperatures.
The linear dependence is no longer present for smaller $Q_2/\alpha$ ratios.
Also, these phenomena seem somewhat robust to adjusting the values of $z$ and $\theta$ away from $\ft43$ and $0$ respectively, but clearly 
a more comprehensive analysis is needed. 
We emphasize once again here, as we did in Section 5, that by neglecting the fluctuations of both gauge fields one obtains an incorrect result
for $\sigma_2^{\rm DC}$, which ignores an important temperature dependent term controlled by $Q_1$.
We leave a more detailed analysis of the temperature dependence of the DC conductivities and a study of the thermal conductivity
(along the lines of \cite{Donos:2014cya,Banks:2015wha}) to future work.

Our model may give rise to a mechanism analogous to that observed in \cite{Sonner:2013aua},
who also examined transport in a gravitational theory with two bulk gauge fields and a dilatonic scalar.
One of the interesting features of the construction of \cite{Sonner:2013aua} is the presence of a finite conductivity
-- specifically, the DC \emph{transconductance} -- without the need to break translational invariance.
Finally, one should keep in mind that a more general understanding of the transport properties in theories such as ours
should take into account extensions of the holographic dictionary to
non-relativistic spacetimes (see \emph{e.g.} \cite{Chemissany:2014xsa,Papadimitriou:2014lia} in the presence of hyperscaling violation).
While we have not attempted to do so here, it is clearly relevant. Moreover, insights from non-relativistic hydrodynamics
might help us understand the role of momentum dissipation in determining the final form of $\sigma^{\rm DC}$.
For instance, an analysis along the lines of \cite{Kiritsis:2015doa} may shed light on the relation between horizon and boundary data, 
and on the interpretation of the latter in our model.

Before closing we should mention that another interesting question in these holographic models is that of the scaling of the Hall angle,
as compared to that of the DC conductivity.
Although here we have not included a magnetic field, by inspecting the structure of the matrix $\sigma_{ij}$
we expect a behavior similar to that observed in \cite{Blake:2014yla}, due to the presence of two scales in our system.
In particular, in analogy with what was seen in \cite{Blake:2014yla} in a different context,
$\sigma_{12}$ scales just like the $\alpha$-dependent parts of $\sigma_{11}$ and $\sigma_{22}$.
Finally, we find it intriguing that (at least in the single charge case $Q_2=0$) the special choice $z=4/3$ discussed by
\cite{Hartnoll:2015sea} also leads to a linear resistivity in our model.
Perhaps more interestingly, in our construction $z=4/3$ is the edge of the range
associated with perturbations that diverge at the non-relativistic boundary.
While the special role played by $z=4/3$ may just be a coincidence, it deserves further attention, as it is of interest to find
explicit gravitational realizations of the scalings singled out in \cite{Hartnoll:2015sea}, and gain insight into their origin.
We leave these questions to future work.

\section*{Acknowledgements}

We are grateful to Aristos Donos, Jerome Gauntlett, Sean Hartnoll and
especially Blaise Gout\'eraux
for comments on the draft.
We also would like to thank Mike Blake for useful conversations.
S.C. would like to thank Cambridge University and the Tokyo Institute of Technology for hospitality during the
final stages of this work.
H-S.L.~is supported in part by NSFC grants No. 11305140, 11375153,
11475148, 11675144 and CSC scholarship No. 201408330017.
The work of H.L.~is supported in part by NSFC grants NO. 11175269, NO. 11475024 and NO. 11235003.
C.N.P.~is supported in part by DOE grant DE-FG02-13ER42020.

\appendix

\section{A General Class of Hyperscaling-violating Solutions}

In this section, we present a class of electrically-charged Lifshitz-Like black branes with hyperscaling violations, carrying magnetic $p$-form fluxes along the brane space.
The Lagrangian consists of the metric, a dilaton, two Maxwell fields and $N$ $p$-form field strengths.
The Lagrangian in general $n$ dimensions is given by
\be
{\cal L} = \sqrt g \Big( R  -\fft12 (\partial \phi)^2 -2\Lambda
e^{\lambda_0  \phi} - \fft14 e^{\lambda_1 \phi} F_{1 \2}^2 -
\fft14 e^{\lambda_2 \phi} F_{2 \2}^2-
\sum_{i=1}^N \fft{e^{\lambda_3 \phi}}{2 p!} {\cal F}^2_{i\,\sst{(p)}}
\Big)\,.
\ee
We consider the ansatz
\bea
&&ds^2 = r^{2 \theta} \big( -r^{2z}\,f\,  dt^2 +
\fft{dr^2}{r^2\, f} + r^2 dx_idx_i \big)  \,,\nn\\
&& \phi = \gamma \, \text{log} \, r  \,, \qquad A_1 = \Phi_1 dt
\,, \quad A_2 = \Phi_2 dt \, , \nn\\
&& {\cal F}^i_{\sst{(p)} } = \alpha \, dx^{(i)}_{j_1}\wedge\ldots
\wedge dx^{(i)}_{j_{p}} \,,
\eea
where $x^{(i)}_j$,\ $1\le j\le p$, denote disjoint sets of transverse-space
coordinates spanning the total $(n-2)$-dimensional transverse space, and so we shall have
\be
Np=n-2\,.
\ee
The equation of motion for the electric field gives
\be
\Phi_1' = Q_1\, r^{z+1-n- \lambda_1 \gamma- (n - 4) \theta} \,, \qquad
\Phi_2' = Q_2\, r^{z+1-n-\lambda_2 \gamma - (n - 4) \theta} \,.
\ee
The solution is
\bea
&& f =1+\frac{\alpha^2  r^{-2 (\theta  p+p+z-1)}}{
 2p (\theta +1)(2 \theta -(n-2p)(\theta +1) +z)} \nn\\
&& \qquad+\frac{Q_2^2 \, r^{-2 (-2 \theta +\theta  n+n+z-3)}}{2 (\theta +1)
(n-2) ((n-2)(\theta+1) + z-2)}- m   r^{2 \theta -(\theta +1) n-z+2}\,,
\eea
with parameters satisfying the relations
\bea
&&\gamma =\sqrt{2 (\theta +1) (n-2) (\theta +z-1)} \,, \quad
\lambda_0=-\frac{\sqrt{2} \theta }{\sqrt{(\theta +1) (n-2)
  (\theta +z-1)}}\,, \nn\\
&&\lambda_1=-\frac{\sqrt{2} ((n-3)\theta +n-2)}{\sqrt{(\theta +1) (n-2) (\theta +z-1)}} \,, \quad
 \lambda_2=-\lambda_3=
\sqrt{\frac{2 (\theta +z-1)}{(\theta +1) (n-2)}}\,, \nn\\
&& Q_1=\sqrt{2 (z-1) ((n-2)(\theta+1) +z)} \,, \nn\\
&&\Lambda=-\ft12 ((n-2)(\theta+1) +z-1) ((n-2 )(\theta+1) +z) \, .
\eea

\end{document}